\documentclass[aps,prd,twocolumn,showpacs,superscriptaddress,groupedaddress,nofootinbib]{revtex4} 

\usepackage[utf8]{inputenc}
\usepackage{graphicx}  
\usepackage{dcolumn}   
\usepackage{bm}        
\usepackage{amssymb,amsmath,diagbox,adjustbox}   
\usepackage{color}
\usepackage{appendix}

\begin{document}

\title[Deep Learning for Cosmological Parameter inference]{Deep Learning for Cosmological Parameter Inference from Dark Matter Halo Density Field}%

\author{Zhiwei Min}
 \affiliation{School of Physics and Astronomy, Sun Yat-Sen University, Zhuhai 519082, China}
 \affiliation{CSST Science Center for the Guangdong–Hong Kong–Macau Greater Bay Area, SYSU, Zhuhai 519082, China}
\author{Xu Xiao}
 \affiliation{School of Physics and Astronomy, Sun Yat-Sen University, Zhuhai 519082, China}
 \affiliation{CSST Science Center for the Guangdong–Hong Kong–Macau Greater Bay Area, SYSU, Zhuhai 519082, China}
\author{Jiacheng Ding}
 \affiliation{School of Physics and Astronomy, Sun Yat-Sen University, Zhuhai 519082, China}
 \affiliation{CSST Science Center for the Guangdong–Hong Kong–Macau Greater Bay Area, SYSU, Zhuhai 519082, China}
 \author{Liang Xiao}
 \affiliation{School of Physics and Astronomy, Sun Yat-Sen University, Zhuhai 519082, China}
 \affiliation{CSST Science Center for the Guangdong–Hong Kong–Macau Greater Bay Area, SYSU, Zhuhai 519082, China}
 \author{Jie Jiang}
 \affiliation{School of Computer Science,South China Normal University,Guangzhou 510631,China}
 \author{Donglin Wu}
 \affiliation{School of Computer Science,South China Normal University,Guangzhou 510631,China}
\author{Qiufan Lin}
\affiliation{Peng Cheng Laboratory, No. 2, Xingke 1st Street, Shenzhen 518000, China}

\author{Yang Wang}
 \affiliation{Peng Cheng Laboratory, No. 2, Xingke 1st Street, Shenzhen 518000, China}

\author{Shuai Liu}
\affiliation{School of Educational Science,Hunan Normal University,Changsha 410081,China}
\author{Zhixin Chen}
\affiliation{School of Educational Science,Hunan Normal University,Changsha 410081,China}
\author{Xiangru Li}
 \affiliation{School of Computer Science,South China Normal University,Guangzhou 510631,China}
\author{Jinqu Zhang}
 \email{zjq@scnu.edu.cn}
 \affiliation{School of Computer Science,South China Normal University,Guangzhou 510631,China}
\author{Le Zhang}
 \email{zhangle7@mail.sysu.edu.cn}
 \affiliation{School of Physics and Astronomy, Sun Yat-Sen University, Zhuhai 519082, China}
 \affiliation{Peng Cheng Laboratory, No. 2, Xingke 1st Street, Shenzhen 518000, China}
 \affiliation{CSST Science Center for the Guangdong–Hong Kong–Macau Greater Bay Area, SYSU, Zhuhai 519082, China}
 \author{Xiao-Dong Li}
 \email{lixiaod25@mail.sysu.edu.cn}
 \affiliation{School of Physics and Astronomy, Sun Yat-Sen University, Zhuhai 519082, China}
 \affiliation{CSST Science Center for the Guangdong–Hong Kong–Macau Greater Bay Area, SYSU, Zhuhai 519082, China}

\date{\today}

\begin{abstract}
We propose a lightweight deep convolutional neural network (lCNN) to estimate cosmological parameters from simulated three-dimensional dark matter (DM) halo distributions and associated statistics.  The training dataset comprises 2000 realizations of a cubic box with a side length of 1000 $h^{-1}{\rm Mpc}$, and interpolated over a cubic grid of $300^3$ voxels, with each simulation produced using  $512^3$ DM particles and $512^3$ neutrinos. Under the flat $\Lambda$CDM model, simulations vary standard six cosmological parameters including $\Omega_m$, $\Omega_b$, $h$, $n_s$, $\sigma_8$, $w$, along with the neutrino mass sum, $M_\nu$.  We find that: 1) within the framework of lCNN, extracting large-scale structure information is more efficient from the halo density field compared to relying on the statistical quantities including the power spectrum, the two-point correlation function, and the coefficients from wavelet scattering transform; 2) combining the halo density field with its Fourier transformed counterpart enhances predictions, while augmenting the training dataset with measured statistics further improves performance; 3) achieving high accuracy in inferring $\Omega_m$, $h$, and $\sigma_8$ by the neural network model, while being inefficient in predicting $\Omega_b$, {  $n_s$}, $M_\nu$ and $w$; 4) { compared to the simple fully connected network trained with three statistical quantities, our CNN yields statistically reduced errors, showing improvements of approximately 23\% for $\Omega_m$, 11\% for $h$, 8\% for $n_s$, and 21\% for $\sigma_8$. Additionally, in comparison with the likelihood-based analysis on $P(k)$ data, our CNN provides much tighter constraints on parameters, especially on $\Omega_m$ and $\sigma_8$.} Our study emphasizes this lCNN-based novel approach in extracting large-scale structure information and estimating cosmological parameters.
 
\end{abstract}

\maketitle


\section{\label{sec:level1}Introduction}

One of the compelling challenges in modern cosmology is the precise estimation of cosmological parameters. With the continuous development of observational techniques, our understanding of the Universe is progressively deepening. However, to comprehensively and accurately understand the evolution and nature of the Universe, key parameters such as the expansion rate and dark energy density need more sophisticated measurement and analysis. This is crucial for validating cosmological models and unlocking the puzzles of the Universe such as  and  Hubble and $S_8$ tensions. High-precision parameter estimates will validate or challenge existing theories, e.g., the $\Lambda {\rm CDM}$ model~\citep{weinberg1989cosmological,peebles2003cosmological,li2011dark}, leading to greater progress in understanding the nature of the Universe.

The large-scale structure (LSS) of the Universe holds significant cosmological information. These vast and intricate formations depict the distribution, accumulation, and evolution of matter in the Universe, serving as crucial observables for comprehending cosmic origins and evolution~\citep{bardeen1986statistics,de1986slice,huchra20122mass,tegmark2004three,guzzo2014vimos}. Through the observation and analysis of LSS, we can track the evolution of the Universe, comprehend its expansion history across various redshifts, explore the formation mechanisms of galaxy clusters and superclusters, and investigate the impacts of DM and dark energy on the evolution of LSS.

At present, the two-point correlation function (2PCF) and its Fourier counterpart, the power spectrum, are the most commonly used statistical tools for analyzing LSS\citep{zhong2024improving}, due to the fact that their sensitivity to both the geometry and the cosmic evolution~\citep{kaiser1987clustering,ballinger1996measuring,eisenstein1998cosmic,blake2003probing,seo2003probing}, allowing for the effective extraction of information regarding Gaussian perturbations. These methods have been successfully applied in analyzing galaxy redshift surveys such as the 2dFGRS \citep{2003astro.ph..6581C}, 6dFGS \citep{Beutler_2011}, the WiggleZ Survey \citep{Riemer_S_rensen_2012}, and the SDSS Survey \citep{york2000sloan,eisenstein2005detection,percival2007measuring,anderson2014clustering,samushia2014clustering,ross2015clustering,beutler2017clustering,sanchez2017clustering,alam2017clustering,chuang2017clustering, Neveux_2020}. However, they encounter difficulties in extracting small-scale information, e.g., $\lesssim 40~h^{-1}{\rm Mpc}$, from the LSS due to the pronounced influence of nonlinear structure evolution caused by gravitational collapse on such scales. Consequently, direct comparisons between observations and theories on the nonlinear scales become challenging.

Alternative statistical measures have been turned to in probing the small-scale properties of the Universe beyond 2PCF. The three-point correlation function~\citep{Sabiu2016Probing,slepian2017large}  has been utilized to improve cosmological constraints, while the more complicated four-point correlation function~\citep{sabiu2019graph}, has demonstrated more stringent constriants. Furthermore, \cite{lavaux2012precision} employed  cosmic voids as a means to probe the cosmic geometry. Additionally, \cite{li2017cosmological} explored the redshift dependence of the 2PCF along the line-of-sight as a probe for cosmological parameters. The symmetry of galaxy pairs was tested~\citep{marinoni2010geometric} and the redshift dependence of the Alcock-Paczynski effect (AP effect) can be exploited to mitigate redshift distortions (RSDs)~\citep{li2015cosmological}. In addition, \cite{li2018galaxy} utilized the tomographic AP method on SDSS galaxy data to obtain a strong constraint on dark energy, and { \cite{Porqueres2021LiftingWL} present a field-level inference on cosmological parameters by analyzing cosmic shear data. }

Recently, the mark weighted correlation function (MCF)~\citep{White_2016} has proposed as an alternative approach. It assigns density weights to various galaxy features to extract non-Gaussian information on LSS. Demonstrated effectiveness in capturing detailed clustering information has led to significantly enhanced constraints on cosmological parameters such as  $\Omega_m$ and $w$~\citep{Yang_2020,lai2023improving,yin2024improving}. Moreover, ~\citep{fang2019,yin2024improving} utilized the $\beta$-skeleton statistics to constrain cosmological parameters.

Although the methods mentioned above can extract rich information from LSS, they also exhibit certain drawbacks. Some methods are overly complex and demand substantial computational resources. In recent years, the rapid development and application of machine learning have introduced new and powerful technical tools for astronomical data analysis, offering innovative solutions to the challenges encountered in survey data analysis~\citep{way2012advances,chen2014data,jordan2015machine,rodriguez2016general,ball2017comprehensive,sen2022astronomical}. Machine learning-based data analysis methods offer significant advantages over traditional approaches in terms of efficiency, accuracy, and feature extraction capabilities. For instance, \cite{Wu_2021,wu2023ai} developed a deep learning technique to infer the non-linear velocity field from the DM density field. In addition, ~\citep{Wang_2023} present a deep-learning technique for reconstructing the dark-matter density field from the redshift-space distribution of dark-matter halos.

\cite{ravanbakhsh2016estimating,pan2020cosmological} utilized convolutional neural networks (CNNs) to extract information from 3D DM distribution and accurately estimate cosmological parameters. Meanwhile, \citep{lazanu2021extracting} employed the Quijote simulation~\citep{villaescusa_Navarro_2020} to estimate cosmological parameters from 3D DM distribution using CNNs, comparing the constraints with those obtained from power-spectrum-based methods. Additionally, \cite{hortua2021constraining} utilized Quijote simulation data to estimate cosmological parameters from a Bayesian neural network, resulting in a posterior distribution of parameters. Recently,~\cite{hwang2023universe} applied the Vision Transformer, known for its advantages in natural language processing, to the estimation of cosmological parameters, and compares its performance with traditional CNNs and 2PCF.

In this study, we explore a deep-learning-based approach to extract cosmological information from the halo number density field. In contrast to previous studies~\citep{lazanu2021extracting,ravanbakhsh2016estimating,pan2020cosmological}, we utilize the halo number density field instead of the DM particle density field. {Recently, \citep{2022OJAp....5E..18M} presented the Graph Information Maximizing Neural Networks, which are capable of quantifying cosmological information from discrete catalog data. In this study, to more realistically reflect real observations, we incorporate several observational effects into our mock samples, such as redshift-space distortion (RSD) effects, coordinate transformations to the fiducial cosmology background, and fixing the halo number density. Meanwhile, the entire parameter space for the fiducial cosmological parameters is jointly inferred, rather than just inferring some of them with other parameters fixed.}  Using the halo catalog of the Quijote's LH$\nu w$ simulation~\citep{villaescusa_Navarro_2020}, our proposed lCNN framework demonstrates the ability to provide reliable constraints on cosmological parameters. Furthermore, we observe that by combining various statistics as input to the lCNN, the performance of the neural network can be noticeably enhanced.

This paper is part of the ``Dark-AI'' project\footnote{https://dark-ai.top/}, a project aims to apply state-of-the-art machine learning algorithms to address frontier problems in cosmology.The structure of this paper is as follows. In Sect.~\ref{sect:2}, we introduce the samples utilized for training and testing, whereas in Sect.~\ref{sect:3}, we outline the architecture of our neural network. Sect.~\ref{sect:4} is dedicated to presenting the results. Finally, we conclude in Sect.~\ref{sect:5} by discussing the results.

\section{Data}\label{sect:2}
To estimate cosmological parameters, training and test samples are constructed using the FoF DM halo catalogues from LH$\nu w$ simulations, a subset of 2000 simulations within the Quijote simulations~\citep{villaescusa_Navarro_2020}--an ensemble of publicly available $N$-body simulations. These simulations utilize the TreePM code Gadget-III~\citep{Springel_2005} and are conducted in boxes with side length $1~h^{-1}{\rm Gpc}$. The LH$\nu w$ simulations offer various cosmological results, evolving $512^3$ DM particles together with $512^3$ neutrino particles. For this study, we focus on the snapshot at $z=0.5$. Beginning from $z=127$, the simulations evolve over time, with matter power spectra and transfer functions obtained from CAMB~\citep{Lewis_2000},  appropriately adjusted. These quantities are used to determine displacements and peculiar velocities via second-order perturbation theory, which are then employed to assign initial particle positions on a regular grid using the 2LPT\cite{Bouchet1994PerturbativeLA}. The simulations are executed by employing Latin-hypercube sampling, a statistical technique for generating a quasi-random sample of parameter values from a multidimensional distribution, with 7 cosmological parameters. The parameter ranges are as follows: $\Omega_m\in [0.1,0.5]$, $\Omega_b\in [0.03,0.07]$, $h\in [0.5,0.9]$, $n_s\in [0.8,1.2]$, $\sigma_8\in [0.6,1.0]$, $M_\nu\in [0,1]~\mathrm{eV}$, and $w\in [-1.3,-0.7]$.
In order to perform a standard likelihood-based analysis using $P(k)$, we also utilize 1000  realizations with different random seeds for a fiducial cosmology to estimate the covariance matrix\cite{villaescusa_Navarro_2020}.The value of the cosmological parameters for the fiducial model are $\Omega_m=0.3175,\Omega_b = 0.049,h=0.6711,n_s = 0.9624,\sigma_8 = 0.834,M_\nu = 0.0~\mathrm{eV},\mathrm{and} \,w=-1$. For this model, the simulatons are run with $1h^{-1}\mathrm{Gpc}$ box and $512^3$ DM particles using 2LPT initial conditions. As a comparison, we also discussed the Rockstar halo catalogs provided by Quijote. The results indicate that the training outcomes of Rockstar are significantly inferior to those of FoF, primarily due to the much lower halo number density of the former. See Appendix \ref{sec:appendixB} for details.

In this study, we performed the following preprocessing steps on the halo catalogs in the Quijote LH$\nu w$ simulations to make it available as data for use by the neural network: 

1) The RSD effect was incorporated along the line of sight (LoS) to more accurately reproduce real observational conditions, as expressed by:
\begin{equation}\label{eq:rsd}
\bm{s}=\bm{r}+\frac{\bm{v} \cdot \hat{z}}{a H(a)}\hat{z},
\end{equation}
where $\bm{r}$, $\bm{s}$ are the position of halos in real space and redshift space respectively. $\hat{z}$ is the unit vector along LoS,  $\bm{v}$ is the peculiar velocity of halos, and $H(a)$ is the Hubble parameter at scale factor $a$. { In preparing each mock, we calculated the RSD effects separately based on the cosmological parameters in each Quijote simulation using Eq.~\ref{eq:rsd}.}

2) Because observational data cannot determine the true cosmological parameters, we must rely on a specific fiducial cosmology. To ensure consistency with the observational data, all mock data must match with this fiducial cosmology background. The fiducial cosmology is derived from Planck 2018 measurements~\citep{Planck:2018vyg}, where $\Omega_m = 0.3071$, $w = -1$. The relation between the Quijote cosmologies and the fiducial cosmology is expressed by:
\begin{equation}\label{eq:fiducial}
s_\perp = s_\perp^0\frac{d^{f}_A(z)}{d_A(z)}\,,~\quad
s_\parallel = s_\parallel^0\frac{H(z)}{H^{f}(z)}
\end{equation}
where $d_A (z)$ and $H(z)$ represent the angular diameter distance and the Hubble parameter at redshift $z$, respectively.  The variables $s_\parallel^0$ and $s_\perp^0$ represent the comoving coordinates in each Quijote simulation. The superscript $f$ denotes the fiducial cosmology, while $\bm{s}_\perp$ and $\bm{s}_\parallel$ represent components perpendicular and parallel to LoS, respectively.  {It should be noted that the analysis employs a fiducial cosmological model to convert redshifts to comoving distances prior to calculating the clustering signal. The objective of this transformation process is to more accurately reflect the data analyzed in real observations~\citep{BOSS:2016wmc}. The results of the analysis are generally not sensitive to the specific parameters of the fiducial cosmology, provided that the fiducial cosmology are not significantly inaccurate.}

3) After conversion from the fiducial cosmology, the box sizes are no longer the same in all three dimensions. Therefore, to conveniently feed the data cubes into the neural network, we cut the converted boxes into sides of equal length, specifically $744~h^{-1}{\rm Mpc}$. Consequently, only the halos within such box in each simulation are considered.

\begin{figure}
	\includegraphics[width=\columnwidth]{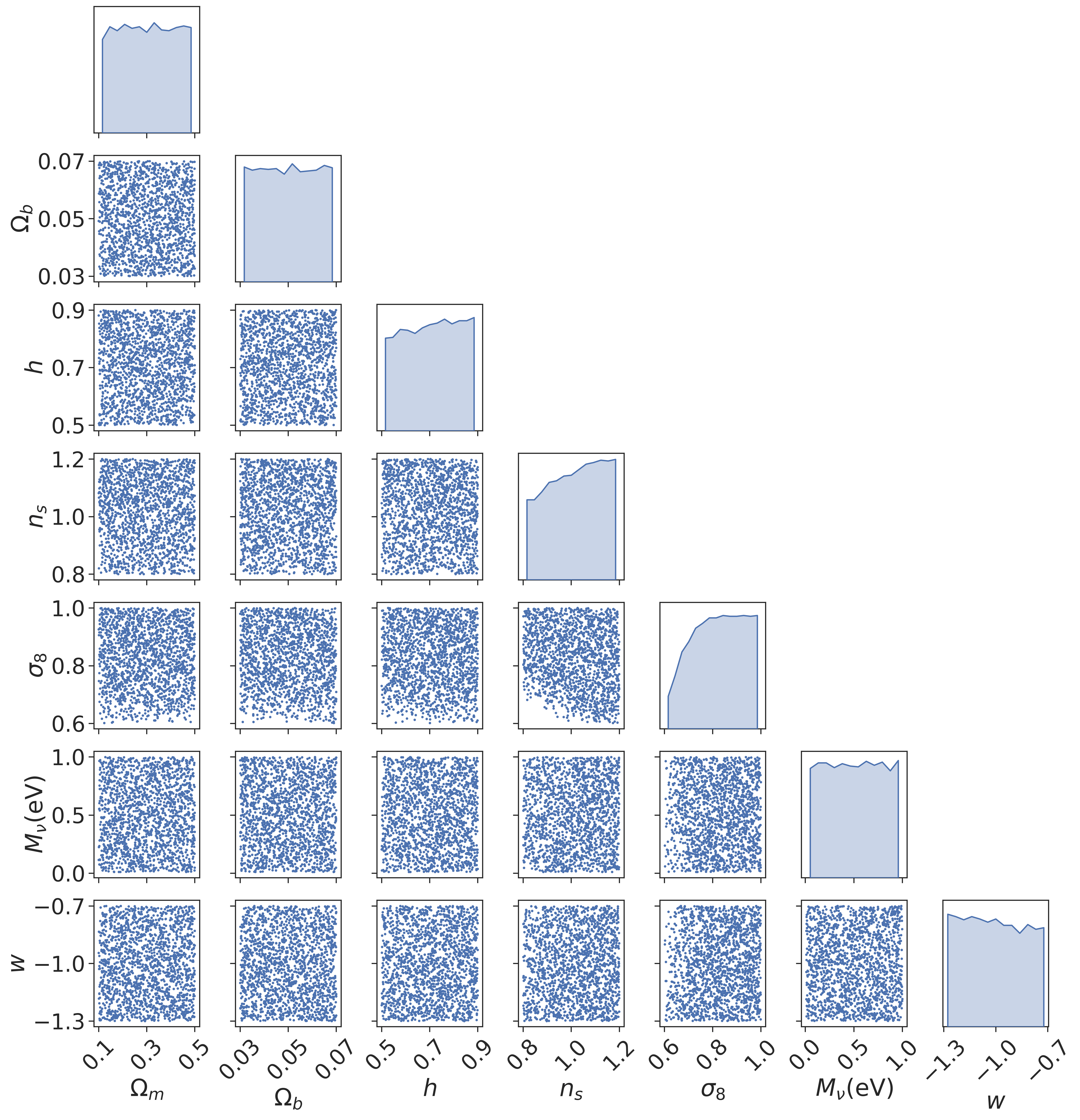}
    \caption{Distribution of cosmological parameters across the simulation boxes following the preprocessing steps on the halo catalogs in the Quijote LH$\nu w$ simulations.As observed, by discarding simulations with halo number densities that are too low, the resulting distributions for each parameter exhibit slight deviations from a uniform distribution.}
    \label{fig:para_dis}
\end{figure}

4) Considering that DM halos with very low mass contribute significant noise, we implemented a cutoff for small mass halos. This cutoff was chosen appropriately such that the number of DM halos has a density equal to $2 \times 10^{-4}~h^3{\rm Mpc}^{-3}$ in each box to be compatible with current spectral observations~{ \cite{Yuan2023UnravelingEL}}. Furthermore, if the halo number density in a simulation box is lower than that value, the box is discarded, resulting in 1710 data cubes remaining. Of these, 1500 are used for training and 210 for testing. Note that, the parameter distributions deviate from a uniform distribution due to the discarding  of some simulation boxes corresponding to different cosmological models, as illustrated in Fig.~\ref{fig:para_dis}. It can be observed that there are noticeably fewer samples with lower $\sigma_8$. This leads to poorer predictive performance in low $\sigma_8$ range. We also discussed replacing the fixed number density of halos with a fixed halo mass cutoff. The results indicate that our results are insensitive to both methods. Regarding the parameter $\sigma_8$, predictions in the lower range have significantly improved. This is because the method of fixing the halo mass cutoff does not require discarding any cosmology, thus maintaining a uniform distribution of parameters. As a result, there are more samples with low $\sigma_8$ values, leading to improved predictive performance. Details are in Appendix \ref{sec:appendixB}.

5) The halo number density field is discretized into mesh cells by assigning the haloes to a $300^3$ mesh using the Cloud-in-Cell (CIC) scheme, with a cell resolution of $(2.48~h^{-1}{\rm Mpc})^3$.

\subsection{Training and Test Samples}\label{sec:Training and Testing Samples}
After preprocessing the simulation data, as mentioned previously, we obtained halo catalogs at the redshift of 0.5 for 1710 cosmological models. We utilized the spatial distribution information of halos together with various associated statistics as both training and test sets for the neural network. This study utilized three datasets for training and testing, as described below.

\begin{figure}
	\includegraphics[width=\columnwidth]{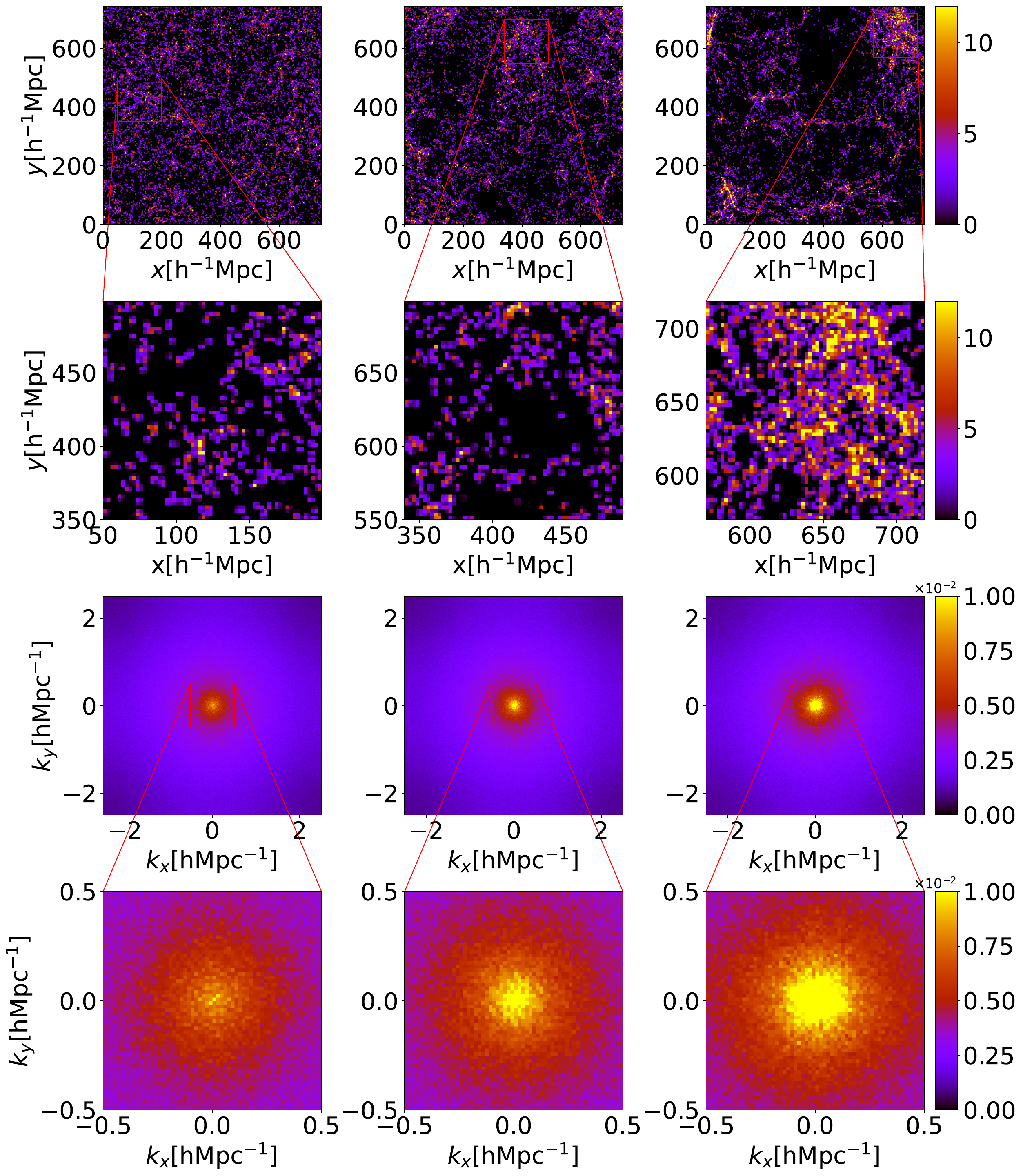}
	\caption{Projected DM halo number density fields of a region of  $744\times 744\times 124~(h^{-1}{\rm Mpc})^3$ and their corresponding Fourier counterparts, selected from the training set. Three different cosmological models are presented from left to right, with the parameters as follows: $\left(\Omega_m, \Omega_b, h, n_s, \sigma_8, M_\nu, w\right)=\{0.32,0.045,0.75,0.93,0.74,0.07{~\mathrm{eV}},-1\}$ (left), $\{0.10,0.049,0.88,1.08,0.93,0.11{~\mathrm{eV}},-0.94\}$ (middle), and $\{0.13,0.054,0.62,0.90,0.94,1.00{~\mathrm{eV}},-1.18\}$ (right). The first and second rows display the spatial distributions of the halo number density field and their zoomed-in plots, where we show the projected field with a thin slice depth of $124~h^{-1} \mathrm{Mpc}$. The third and last rows correspond to the amplitudes of the corresponding Fourier modes of the density fields and their zoomed-in versions, where the depth along LoS is within $\Delta k \in[-0.5,0.5] ~h\mathrm{Mpc}^{-1}$.     
 }
	\label{fig:density field}
\end{figure}

\textbf{Dataset A}: we utilized the three-dimensional distribution of the DM halo number density field, $n(\bm{x})$, which is interpolated onto a $300^3$ mesh with a resolution of $(2.48~h^{-1}{\rm Mpc})$ along each side, to extract the input cosmological parameters. The first and second rows of Fig.~\ref{fig:density field} display the projected halo number density fields in three different cosmological models, along with their corresponding zoomed-in plots.

\textbf{Dataset B}: We utilized the Fourier-transformed halo density field with $300^3$ grids. Letting $\tilde{\delta}(\mathbf{k})$ denote the Fourier transform of the overdensity $\delta(\mathbf{x})$, defined by
\begin{equation}\label{eq:fourier transform}
	\tilde{\delta} (\bm{k}) = \int \frac{d^3 x}{(2\pi)^{3/2}} 
 \delta(\bm{x})\exp(-i\bm{k}\cdot \bm{x})\,,
\end{equation}
where $\delta(\bm{x}) = n(\bm{x})/\bar{n}-1$ is the density contrast, a dimensionless measure of overdensity at each point. 

In practice, to complement $n(\bm{x})$, we retain only the low-frequency (i.e. large-scale) modes in the Fourier space field, which are not captured by the configuration space field. Specifically, we filter $\tilde{\delta}(\bm{k})$ with $|k| < 0.5 h\mathrm{Mpc}^{-1}$, resulting in a datacube of Fourier modes on a $60^3$ grid. The third and fourth rows of Fig.~\ref{fig:density field} show the amplitudes of the Fourier fields for the three different cosmological models,  along with their corresponding zoomed-in plots. Here, the zero-frequency mode is located at the center of each plot. Note that both amplitude and phase are input into the neural network, where each Fourier mode can be expressed as $\tilde{\delta} = Ae^{i\phi}$, with $A$ representing amplitude and $\phi$ representing phase.

\textbf{Dataset C}: in addition to the density field information, we have integrated various statistics into our training samples. These statistics comprise the two-point correlation function of halos, $\xi(r)$, and the corresponding power spectrum, $P(k)$, and the wavelet scattering transform $(\rm{WST})$ coefficients, labeled as $S_n$, where $n$ denotes the order of WST coefficients.

The power spectrum is given by the following average over Fourier space:

\begin{equation}\label{eq:pk}
\left<\tilde{\delta}(\bm{k})\tilde{\delta}^*(\bm{k}')\right>=(2\pi)^3 P(k)\delta^3(\bm{k}-\bm{k}').\end{equation}

The relationship between $P(k)$ and $\xi(r)$ is a Fourier transform, which can be mathematically expressed as follows,

\begin{equation}
\xi(r)\equiv \left<\delta (\bm{x})\delta (\bm{x-r})\right> = \int \frac{d^3 k}{(2 \pi)^3} P(k) e^{i \bm{k} \cdot\bm{r}}\,.
\label{eq:pcf}
\end{equation}

Considering the relatively large uncertainty of statistics at small scales due to noise, $\xi(r)$ and $P(k)$ are normalized by their mean value,namely we only utilized their shapes and focused on specific ranges: $r\in \left[20, 200\right]~h^{-1}{\rm Mpc}$ for $\xi(r)$ and $k\in \left[0.05, 0.6\right]~h{\rm Mpc^{-1}}$ for $P(k)$. In other words, we discarded magnitude information, keeping only shape information.

\begin{figure}
	\includegraphics[width=\columnwidth]{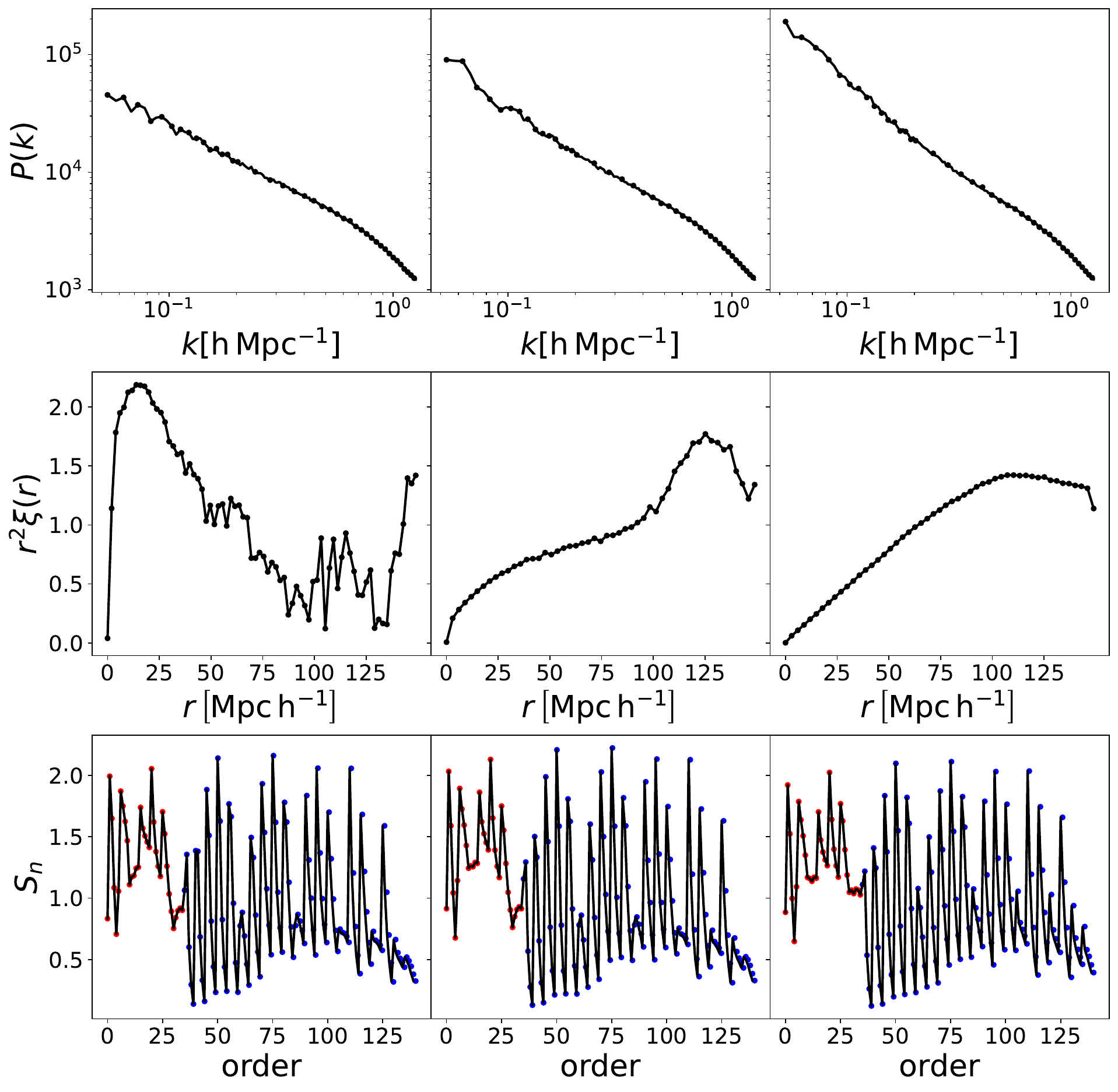}
	\caption{Three statistics are utilized as training data for cosmological parameter inference. They are derived from the same density fields representing three distinct cosmologies, as depicted in Fig.~\ref{fig:density field}. According to Eqs.~\ref{eq:pk},~\ref{eq:pcf} and~\ref{eq:wst}, the power spectra, 2PCFs, and WST coefficients for the halo number density fields are presented from top to bottom, respectively. In the third row, the red and blue dots represent the WST coefficients of $S_1$ and $S_2$, respectively, each of which is normalized by its mean value.}
	\label{fig:statistics}
\end{figure}
The wavelet scattering transform (WST) was originally introduced in the context of signal processing in computer vision, as discussed by~\cite{wst1,mallat2012group}. This method serves the purpose of capturing the statistical properties inherent in an input field. In the WST framework, an input field $I(\bm{x})$ undergoes two primary nonlinear operations: wavelet convolutions and modulus calculations. Essentially, when $\Psi_{j_1, l_1}(\bm{x})$ denotes an oriented wavelet probing a scale $j_1$ and angle $l_1$, the WST operation transforms $I(\bm{x})$ as follows:

\begin{equation}
I'(\bm{x}) =\left|I(\bm{x}) \otimes \Psi_{j_1, l_1}(x)\right|\,.
\label{eq:wst}
\end{equation}
Here, $\otimes$ represents convolution. The averaging of this operation produces a WST coefficient $S_n$, essentially a real number describing the characteristics of the field. Through the utilization of a set of localized wavelets $\Psi_{j_1, l_1}(\bm{x})$, exploring different scales $j_1$ and angles $l_1$, repeated iterations of this process generate a scattering network. The WST coefficients, $S_n$, up to order $n=2$, are determined by the following relationships:

\begin{equation}
\begin{aligned}
S_0 & =\left<|I(\bm{x})|\right>, \\
S_1\left(j_1, l_1\right) & =\left<\left|I(\bm{x}) \otimes \Psi_{j_1, l_1}(\bm{x})\right|\right>, \\
S_2\left(j_2, l_2, j_1, l_1\right) & =\left<\left|\left(\left|I(\bm{x}) \otimes\Psi_{j_1, l_1}(\bm{x})\right|\right) \otimes \Psi_{j_2, l_2}(\bm{x})\right|\right>\,.
\end{aligned}\label{eq:wst_coeff}
\end{equation}
Here, $<\cdot>$ denotes averaging over samples. Generally, a family of wavelets $\Psi_{j_1, l_1}(\bm{x})$ can be generated by applying dilations and rotations to a mother wavelet. In our study, the mother wavelet is a solid harmonic multiplied by a Gaussian envelope, taking the form of
\begin{equation}
    \Psi_l^m(\bm{x}) = \frac{1}{(2\pi)^{3/2}}\mathrm{e}^{-|\bm{x}|^2/2\sigma^2}|\bm{x}|^{l}Y_l^m(\frac{\bm{x}}{\left|\bm{x}\right|})\,,
\end{equation}
where, $Y^m_l$ represents the Laplacian spherical harmonics, and $\sigma$ denotes the Gaussian width measured in field pixels. In this study, we set $\sigma = 0.25$. Given a 3D input field, along with a total number of spatial dyadic scales $J$ and total orientations $L$, WST coefficients can be calculated to any order. Here, the coefficient order is defined as a function of $(j, l)$, where $j \in [0, 1, ..., J-1, J]$ and $l \in [0, 1, ..., L-1, L]$. Detailed information on the coefficients can be found in~\cite{valogiannis2023precise}. 
In our analysis, we set $J=6$ and $L=4$, resulting in a total of 140 WST coefficients, excluding $S_0$. In summary, the WST coefficients in our work are 
\begin{equation}
    \begin{split}
        S_0& = \left<|I(\bm{x})^q|\right>,\\
        S_1(j_1,l_1) &=\left<\big(\sum_{m=-l_1}^{m=l_1}|I(\bm{x})\otimes\Psi_{j_1,l_1}^m (\bm{x})|^2\big)^{q/2}\right>,\\
        S_2(j_2,j_1,l_1) &=\left<\big(\sum_{m=-l_1}^{m=l_1}|U_1(j_1,l_1)(\bm{x})\otimes\Psi_{j_2,l_1}^m (\bm{x})|^2\big)^{q/2}\right>\,,
    \end{split}
    \label{eq:wst coeff}
\end{equation}
with 
\begin{equation}
    U_1(j_1,l_1)(\bm{x}) = \sum_{m=-l_1}^{m=l_1}|I(\mathbf{x})\otimes\Psi_{j_1,l_1}^m (\bm{x})|^2\,,
\end{equation}
where $q$ is a specified power governing operations on a target field. Choosing $q>1$ or $q<1$ highlights overdense or underdense regions, respectively, while $q=1$ represents the basic WST scenario. In our analysis, we consider all three cases: $q=0.5$, $q=1$, and $q=2$. Fig.~\ref{fig:statistics} presents these three statistics, derived from the same three simulation boxes as depicted in Fig.~\ref{fig:density field}. For WST coefficients, only $q=0.5$ is displayed.

Finally, employing the principal component analysis (PCA) technique, an efficient compression scheme is utilized to retain most of the signal information encoded in the data while projecting out the noise-dominated modes. The original $\xi(r)$ has 266 bins, $P(k)$ has 243 bins, and  $S_1$ , $S_2$ totally consist of 420 bins. Through PCA, each measurement statistic is compressed into a one-dimensional vector of 20 dimensions.

\section{Method}\label{sect:3}
\begin{figure*}
	\includegraphics[width=1.9\columnwidth]{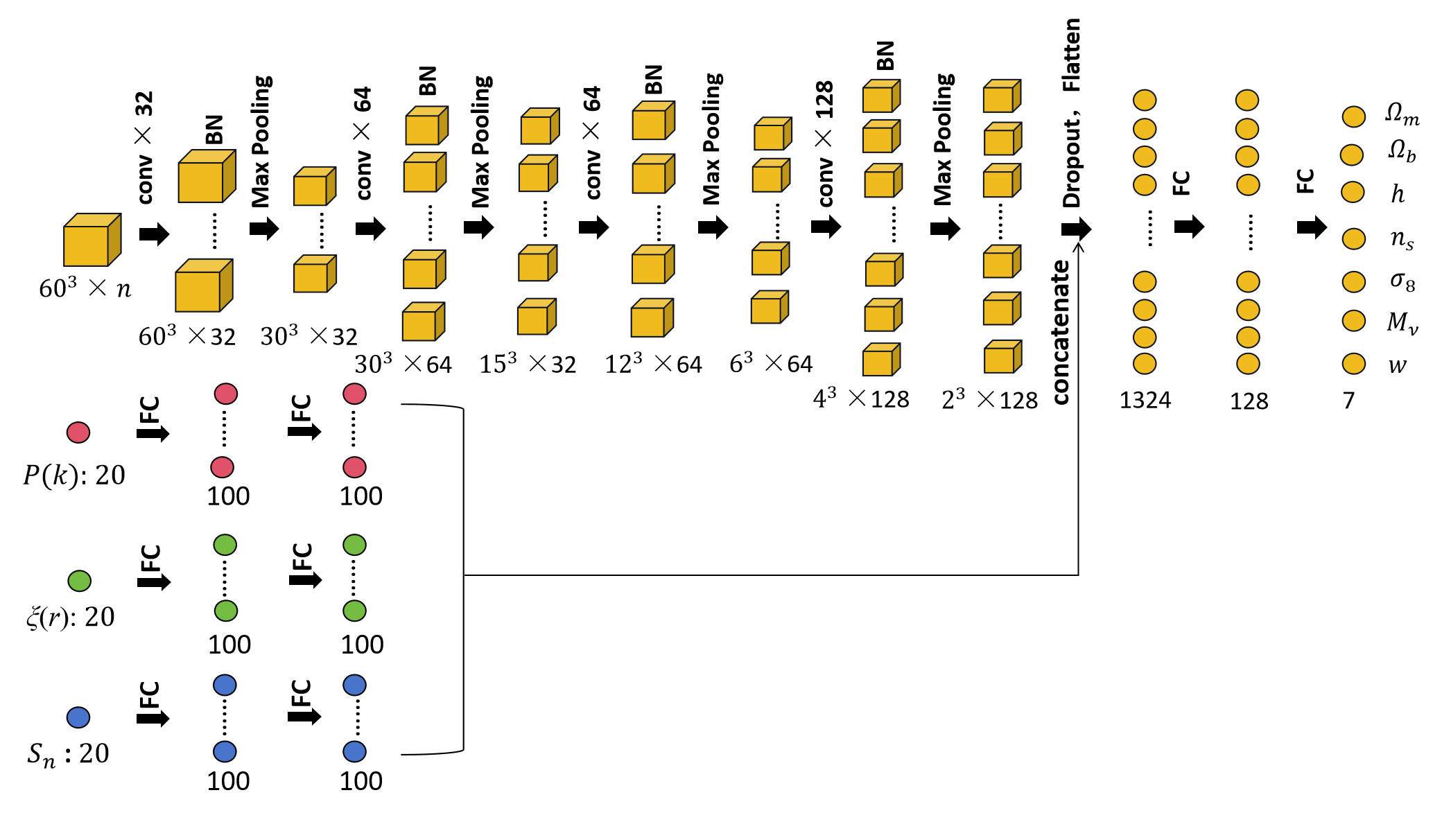}
	\caption{ Architecture of the proposed lightweight deep convolutional neural network (lCNN) for parameter estimation. Starting with a cube of size $60^3\times n$, where $n=1$ for the density field and $n=3$ for the combination of density and Fourier modes (2 for amplitude and phase), the network comprises several convolutional layers. Each convolutional layer is followed by Batch Normalisation (BN) layers to enhance convergence, and Max Pooling layers to reduce dimensionality. The dropout technique is also used for preventing the network from overfitting. Subsequently, the network is flattened and includes two fully connected layers consisting of 1024 and 128 neurons. The output layer with seven neurons corresponds to the original input parameters $\{\Omega_m,\Omega_b, h, n_s, \sigma_8,  M_\nu,w\}$.  For feature extraction, four convolutional layers with filters 32, 64, 64, and 128 are applied. If the statistical measurements, including the power spectrum, 2PCF, and WST coefficients, each having a dimension of 20, are utilized, we then construct two fully connected layers to transform the dimension of each statistic to 100. These layers are then concatenated with the density field network.}
	\label{fig:cnn}
\end{figure*}

To fully exploit the three-dimensional nature of the data, we employed a deep 3D convolutional network. After investigating several architectures, we propose a lightweight deep convolutional neural network (lCNN) that is efficient and present high performance in parameter estimation. Fig.~\ref{fig:cnn} schematically depicts lCNN designed for determining cosmological parameters.

The network comprises three types of layers: 3D convolutions which is followed by batch normalization, max pooling layers,and fully connected layers. It begins with a $60^3$-voxel input layer representing the density field. When incorporating the Fourier transform of the density, two extra channels are introduced to accommodate the amplitude and phase of the Fourier modes. Thus, we represent the dimension of the input data cube as $60^3 \times n$, where $n$ is for the number of channels utilized. Following this 4 convolutional layers are applied, and each is accompanied  by batch normalization and a max-pooling layer with a kernel $(2,2,2)$ for dimensionality  reduction. The size of kernel in convolutional layers is $(3,3,3)$ except the third, which is $(4,4,4)$. { Specifically, in the first two convolutional layers, we performed padding operations, which involve adding extra layers of zeros around the input data matrix before applying the convolution operation. The main purpose of padding is to preserve the spatial dimensions of the input volume~\cite{Li2020ASO}}. {For comparison, we examined circular padding, which is more suitable for data with periodic boundary conditions. We found that the results did not show significant changes, as 97.7\% to 98.1\% of the pixel values in the halo density field are zero. For details, see Appendix \ref{sec:appendixB}.} After four 3D convolutions,the input data information are encoded by  $128\times 2^3$ voxels, which are transited into a standard deep neural network after the flatten operation. Here, we introduce two new hidden layers, with 1324 and 128 neurons respectively, before concluding with a seven-neuron output layer. This output layer corresponds to the seven parameters that have been varied in the simulations. Additionally, when employing statistical measurements such as power spectra, 2PCF, and WST coefficients, each measurement originally has a dimension of 20. We then construct two fully connected layers to transform the dimension of each statistic to 100, which are concatenated with the output features of lCNN before passing them through the fully connected layers. {It is worth mentioning that changing the number of convolutional layers and fully connected layers in the CNN does not have a significant impact on our results. This indicates that our results exhibit a certain robustness to the choice of neural network architecture. For specific details on this discussion, please refer to Appendix \ref{sec:appendixA}.}

Throughout the network, rectified linear unit (ReLU) activation functions are employed. For optimization, the Adam optimizer~\citep{kingma2017adam} is utilized with a learning rate of $5 \times 10^{-5}$. For our machine learning task, we opted for the widely-used Mean Squared Error (MSE) loss function. This metric quantifies the average squared difference between predicted and true values, defined as

\begin{equation}
    {\rm MSE} = \frac{1}{N}\sum^{N}_{i=1}\left(y^{\rm pred}_i - y^{\rm true}_{i}\right)^2
    \label{eq:mse}
\end{equation}

In the training and testing process, we employed a density field that has been interpolated into a $300^3$ grid using the CIC scheme, as previously mentioned. For training purposes, we divided a single $300^3$ data cube into $5^3$ sub-boxes, each with dimensions of $60^3$. When combining the density field in Fourier space, which has dimensions of $60^3 \times 2$, the density and its Fourier transform were concatenated into a sub-box with dimensions of $60^3 \times n$, where $n=3$. When utilizing only statistical measurements without incorporating density fields, parameter estimation is exclusively performed using the random forest network.

\begin{figure}
	\includegraphics[width=\columnwidth]{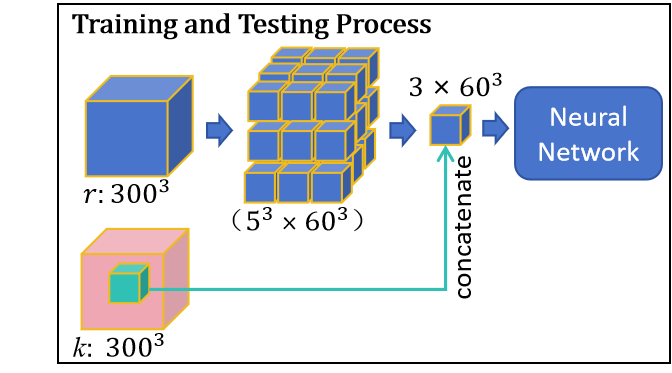}
	\caption{Training and testing processes. Data cubes are divided into $5^3$ sub-boxes of dimensions $60^3$. When incorporating density fields in Fourier space, with dimensions $60^3 \times 2$, they are concatenated into sub-boxes of size $60^3 \times 3$.We combine the density fields in configuration and Fourier spaces to utilize clustering information from both small-scale and large-scale structures. During training, sub-boxes are randomly selected for each epoch, while testing involves feeding all $5^3$ sub-boxes into the neural network, with predictions averaged to estimate the cosmological model.}
	\label{fig:training and testing}
\end{figure}

During training, for each epoch, we randomly selected a sub-box to feed into the neural network. For testing, all $5^3$ sub-boxes are fed into the neural network, and the predictions are averaged to estimate the cosmological model. The training and testing processes are schematically depicted in Fig.~\ref{fig:training and testing}.

\subsection{Evaluation Metrics}
Once the data is divided into training and test sets, we proceed to estimate the cosmological parameters for inputs outlined in subsequent sections. We evaluate the performance of each model through four approaches on the test set to quantify the results: 1) plotting the predicted values against the ground truth for the test set, quantified by the coefficient of determination $R^2$,$R^2$ ranges from $0$ to $1$, where $1$ represents perfect inference.; 2) calculating the averaged bias (Bias) for each parameter; 3) calculating the relative squared error (RSE) for each parameter; 4)  calculating the Root  Mean Square (RMSE) for each parameter. These quantities are defined as follows:

\begin{equation}\label{eq:r2}
R^2=1-\frac{\sum_i\big(y_i^{\rm pred }-y_i^{\rm true }\big)^2}{\sum_i\big(y_i^{\rm true }-\bar{y}^{\rm true }\big)^2}\,,
\end{equation}

\begin{equation}
    {\rm Bias} = \frac{1}{N}\sum_{i}\big(y_{i}^{\rm pred} - y_{i}^{\rm true}\big)\,,
    \label{eq:bias}
\end{equation} 

\begin{equation}
    {\rm RSE} = \frac{\sum_{i}\big(y_{i}^{\rm pred}-y_{i}^{\rm true}\big)^2}{\sum_{i}\big(y_{i}^{\rm true}-\bar{y}^{\rm true}\big)^2}\,,
    \label{eq:rse}
\end{equation}
and
\begin{equation}\label{eq:rmse}
{\rm RMSE}=\sqrt{\frac{1}{N}\sum_{i}\big(y_i^{\rm pred}-y_i^{\rm true }\big)^2}\,,
\end{equation}
where the summation runs through the entire $N$ test samples, with the bar indicating the average.  The $R^2$ quantifies the fraction by which the error variance is less than the true variance, while the RMSE provides an overall measure of the model's prediction accuracy, with lower values indicating better performance.  Similarly, RSE measures the relative error between predicted and true values by comparing the squared difference between them. On other hand, Bias denotes the systematic error of predictions from true values. A Bias close to zero indicates that, on average, the model is making predictions that are unbiased.

\section{Results}\label{sect:4}
In this section, we present the results obtained from various models using different inputs, including the density field, its Fourier modes, the three statistical measurements, and their combinations. Additionally, we compare the performance of predictions made by the random forest model using statistical measurements alone.

Five distinct models were devised to evaluate the optimal choice among different datasets as inputs, including:
\begin{enumerate}
\item Model ``${\rm CNN}(r)$'': utilizing solely the density field. 
\item Model ``CNN ($r+k$)'': incorporating the density field along with its Fourier modes.
\item Model ``CNN($r$)+FC(statistics)'': employing the density field together with three statistics including $P(k)$, $\xi(r)$, and $S_n$.
\item Model ``CNN($r+k$)+FC(statistics)'': combining the case ``CNN ($r+k$)'' with three statistics.
\item Model ``RF(statistics)'': using only the three measured statistics.
\end{enumerate}

{ Here, ``CNN'' and ``FC'' represent the convolutional layers and fully connected layers of lCNN, while "RF" represents the random forest network. For comparison, the model ``FC(statistics)'' incorporating the three statistics was trained using a fully connected network similar to the FC layers of our CNN, and ``FC(Pk)'' corresponds to the FC network trained using the statistics of the power spectrum alone.}

\begin{figure}    
	\includegraphics[width=\columnwidth]{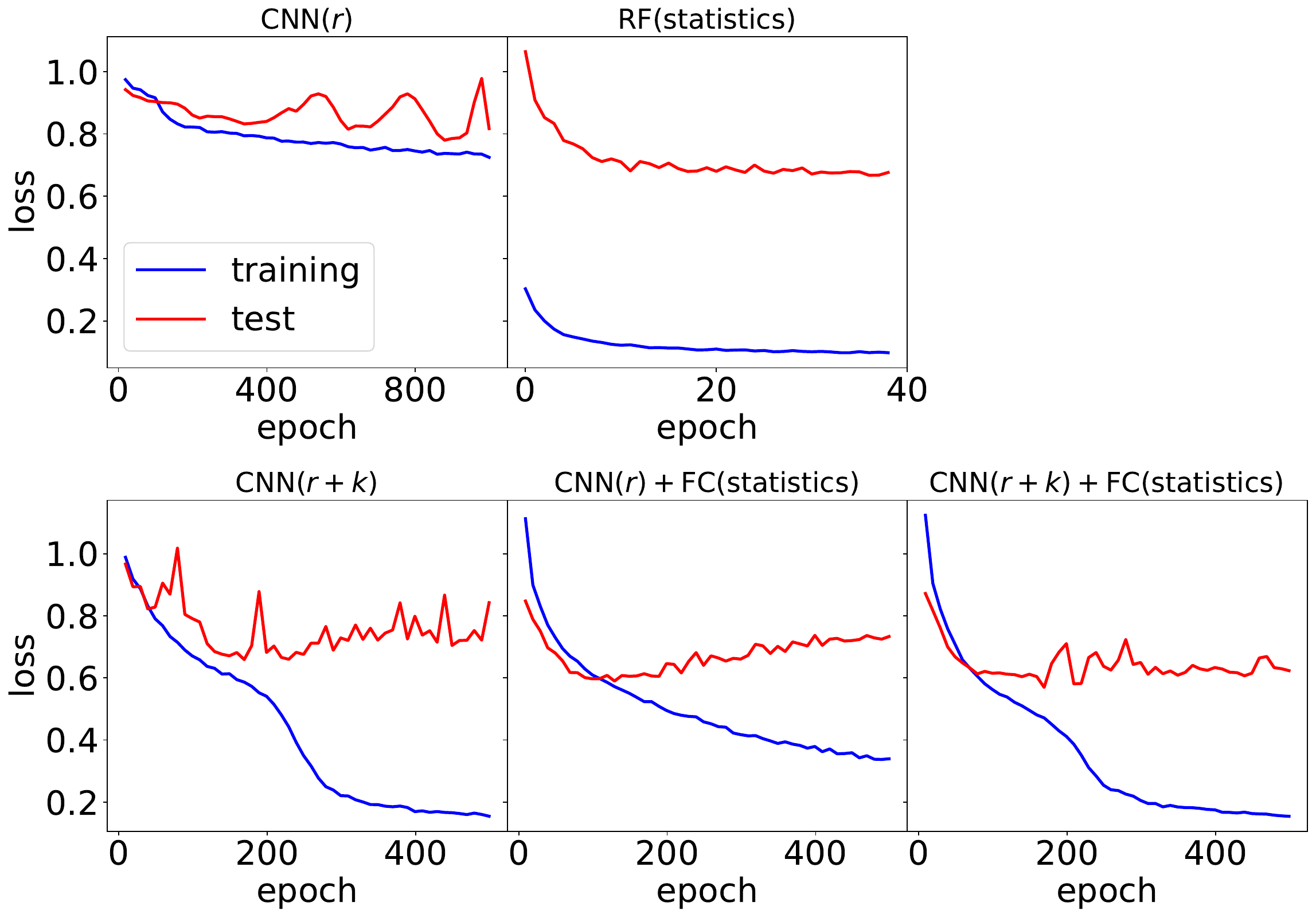}
	\caption{Loss functions, as defined in Eq.~\ref{eq:mse}, for five different cases for both the training and testing datasets across epochs. The blue and red curves represent training and testing losses, respectively, covering 87.7\% and 12.3\% of the full dataset, respectively. When using ``CNN($r$)'' alone, the training loss decreases gradually, converging to $0.77$. 
  For ``CNN($r+k$)+FC(statistics)'', the loss reaches its minimum approximately at epoch 200, outperforming all other cases. Such case also yields the lowest testing loss values, demonstrating best performance for parameter estimation.}
	\label{fig:loss}
\end{figure}

Fig.~\ref{fig:loss} displays loss curves against the number of epochs for the five different training sets. The blue and red curves represent the loss for the training and testing data sets, respectively, corresponding to 87.7\% and 12.3\% of the full dataset.

It can be observed that when using ``CNN($r$)'' alone, the loss function on the training set decreases slower compared to other cases, gradually converging to $0.77$. { Incorporating $k$-space data or statistics has the effect of reducing loss on the test data, which is an indication of a modest improvement in performance. However, on the training data, there is a considerable reduction in loss, which suggests that the model has achieved its optimum generalisation point in the test set. In such cases, the lowest test set loss is typically taken to indicate the optimal performance of the model.}

For the case of ``CNN($r+k$)+FC(statistics)'', the loss function drops to the lowest value of 0.57 at epoch about 200, outperforming all other cases.  Importantly, comparing with the loss functions for the testing dataset, combining both fields and statistics results in the lowest loss values. Thus, as expected, this case demonstrates the best performance for parameter estimation, as we will demonstrate later. Additionally, we trained lCNN until achieving the lowest loss values for the test set. 

\begin{figure*}
	\includegraphics[width=1.6\columnwidth]{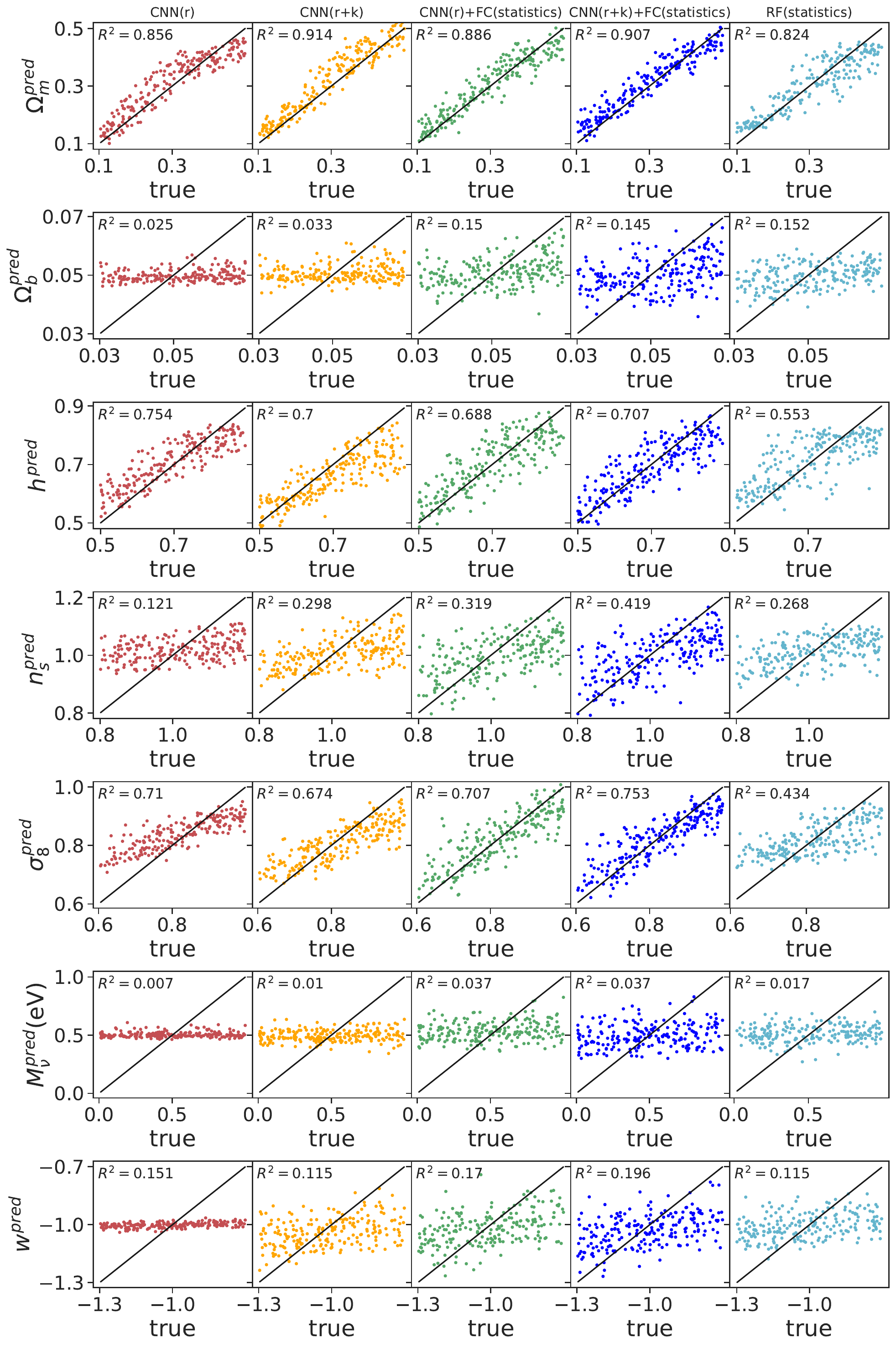}
	\caption{Comparison between the true values of $(\Omega_m, \Omega_b, h, n_s, \sigma_8, M_\nu, w)$ (from top to bottom) from the test sample and their predictions from the five models (from left to right) of lCNN. The black lines indicate perfect prediction matching the truth. The resulting $R^2$ values for each case are also shown. On average, the model ``CNN($r+k$)+FC(statistics)'' exhibits better performance for prediction than other models. lCNN performs well for predicting parameters $\Omega_m$, $h$, and $\sigma_8$, but the prediction ability becomes weaker for $\Omega_b$, $M_\nu$, $n_s$, and $w$. For comparison, predictions from the random forest (the fifth column) trained with the three statistics only, including the power spectrum, the 2PCF, and the WST coefficients, are shown.}
	\label{fig:prediction}
\end{figure*}

In Fig.~\ref{fig:prediction}, we present the actual predictions from our designed lCNN using the entire test dataset. The corresponding $R^2$ value for each case is also listed in each panel. For comparison, black curves are drawn to represent perfect parameter recovery ($R^2=1$), where the correlation between the predicted and true values is 100\%.

The five panels in each row correspond to the five input models for fixed cosmological parameters. Results for the seven cosmological parameters are presented from top to bottom, respectively. The prediction accuracies for $\Omega_m$, $\sigma_8$ and $h$ from lCNN are significantly higher compared to other parameters. Especially, the predicted $\Omega_m$ values closely match the ground truth (black lines), with relatively small scatters. As expected, overall, the model ``CNN($r+k$)+FC(statistics)'' demonstrates the best performance for parameter predictions among other models, evident from its highest average $R^2$ value. However, none of the five models perform well for predicting $\Omega_b$, $M_\nu$, and $w$, with the highest $R^2$ values only reaching $0.152$, $0.037$, and $0.196$, respectively. This is because these parameters do not visibly imprint unique features in LSS in our simulation mocks and also degenerate with other parameters.

Additionally, LSS is sensitive to the total density of $\Omega_m$, rather than the relatively small quantity of $\Omega_b$, as no baryonic feedback is considered in the cold DM simulations. Moreover, one snapshot at $z=0.5$ cannot effectively distinguish the different dark energy equations of state $w$. Since both lCNN and the random forest fail to provide effective predictions for $\Omega_b$, $M_\nu$ and $w$, we do not display their results in the following.

\begin{table*}
\begin{adjustbox}{width=2\columnwidth,center}
\centering
\begin{tabular}{c|c|c|c|c|c|c}
\hline
\hline
\diagbox{Parameter}{Bias/RMSE/RSE}{Model}
 &   CNN($r$)& CNN($r+k$) & CNN($r$)+FC(statistics) & CNN($r+k$)+FC(statistics)&RF (statistics)& { FC(statistics)} \\ \hline
 $\Omega_m$ &  0.013/0.047/0.161& 0.023/0.041/0.126 & 0.005/0.040/0.118 & 0.014/0.039/0.113& -0.006/0.050/0.187&-0.009/0.050/0.182 \\ \hline
 $h$ & 0.011/0.060/0.284 & -0.039/0.073/0.429 & 0.004/0.063/0.314 & -0.004/0.061/0.295&0.008  /0.077/0.454&0.004/0.067/0.361\\ \hline
 $n_s$ &  0.004/0.110/0.881& -0.009/0.099/0.717 & -0.020/0.099/0.713 & -0.020/0.091/0.610&-0.003/  0.099/0.740&-0.011/0.095/0.657\\  \hline
 $\sigma_8$ & 0.017/0.072/0.440& -0.014/0.067/0.388 & -0.0007/0.059/0.610 & -0.001/0.059/0.293 &-0.0003/  0.078/0.582&0.007/0.069/0.396\\ \hline
\end{tabular}
\end{adjustbox}
\caption{Summary of the measured evaluation metrics of Bias, RMSE, and RSE for the { six} models across four cosmological parameters. A small value for Bias, RMSE, or RSE indicates that the predictions made by the model are close to the true values, suggesting that the performance is relatively good.}
\label{tab:bias_rmse}
\end{table*}

In Tab.~\ref{tab:bias_rmse}, the Bias, RMSE, and RSE metrics are presented for detailed comparison across the four cosmological parameters among the five models. The results agree with the $R^2$ values depicted in Fig.~\ref{fig:prediction}. The ``CNN($r+k$)+FC(statistics)'' model demonstrates the smallest RMSE and RSE values across almost all of these parameters, indicating high accuracy and small uncertainty estimation compared to other models. Notably, Bias values for all models closely match the true values within a $2\sigma$ level compared to RMSE, highlighting the robustness of the networks and negligible systematic errors.

\begin{figure}
	\includegraphics[width=\columnwidth]{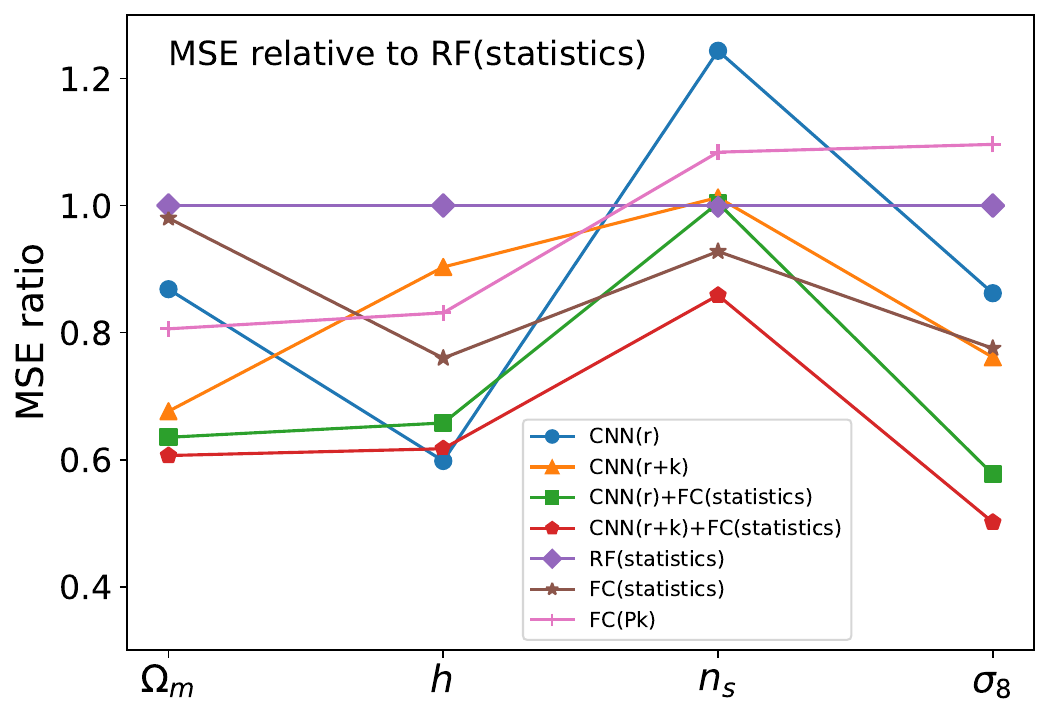}
	\caption{RMS values for different models  relative to those of the model ``RF(statistics)'' using the simple random forest network across cosmological parameters, including $\Omega_m$, $h$, $n_s$, and $\sigma_8$. The smaller the value of MSE indicates better performance in parameter prediction. {The models ``FC(statistics)'' and ``FC(Pk)'' are trained using a fully connected network similar to the FC layers of our CNN, incorporating all statistics and $P(k)$ respectively.}.Therefore, the model ``CNN($r+k$)+FC(statistics)'' (red) exhibits the best performance among other models. As the predictions for $\Omega_b$,$M_\nu$  and $w$ are ineffective by lCNN, we do not show their results. }
	\label{fig:mserela}
\end{figure}

To emphasize the MSE values for different models, in Fig.~\ref{fig:mserela}, we display the relative MSE values compared to the model ``RF(statistics)''. This is represented as the ratio of MSE for each model to that from the random forest network. Since our trained lCNN models are ineffective for $\Omega_b$, $M_\nu$, and $w$, due to the much low $R^2$ values, we only compare MSE values of the four parameters ($\Omega_m$, $h$, $n_s$, $\sigma_8$) relative to that of model ``RF(statistics)''. From this comparison, we observe that, except for the parameter $n_s$ with the model ``CNN($r$)'', lCNN performs significantly better than with ``RF(statistics)''. {However, compared to random forests (RF), when training on statistical quantities using the same fully connected layers as our lCNN, denoted as the model ``FC(statistics)'', there was an improvement in the performance of parameters $n_s$, $h$ and $\sigma_8$. }Additionally, the combination of the density field and the Fourier modes, i.e., the model ``CNN($r+k$)'', performs better than using the density field alone, corresponding to the model ``CNN($r$)'', except for the parameter $h$. Moreover, feeding the three measured statistics to lCNN further enhances the accuracy of prediction, effectively lowering the MSE values. The best performance is achieved for the model ``CNN($r+k$)+FC(statistics)'' (red line), reducing MSE by about {5--37}\% when compared with {``FC(statistics)''} model across the parameters.

\begin{figure*}[htpb]
	\includegraphics[width=1.6\columnwidth]{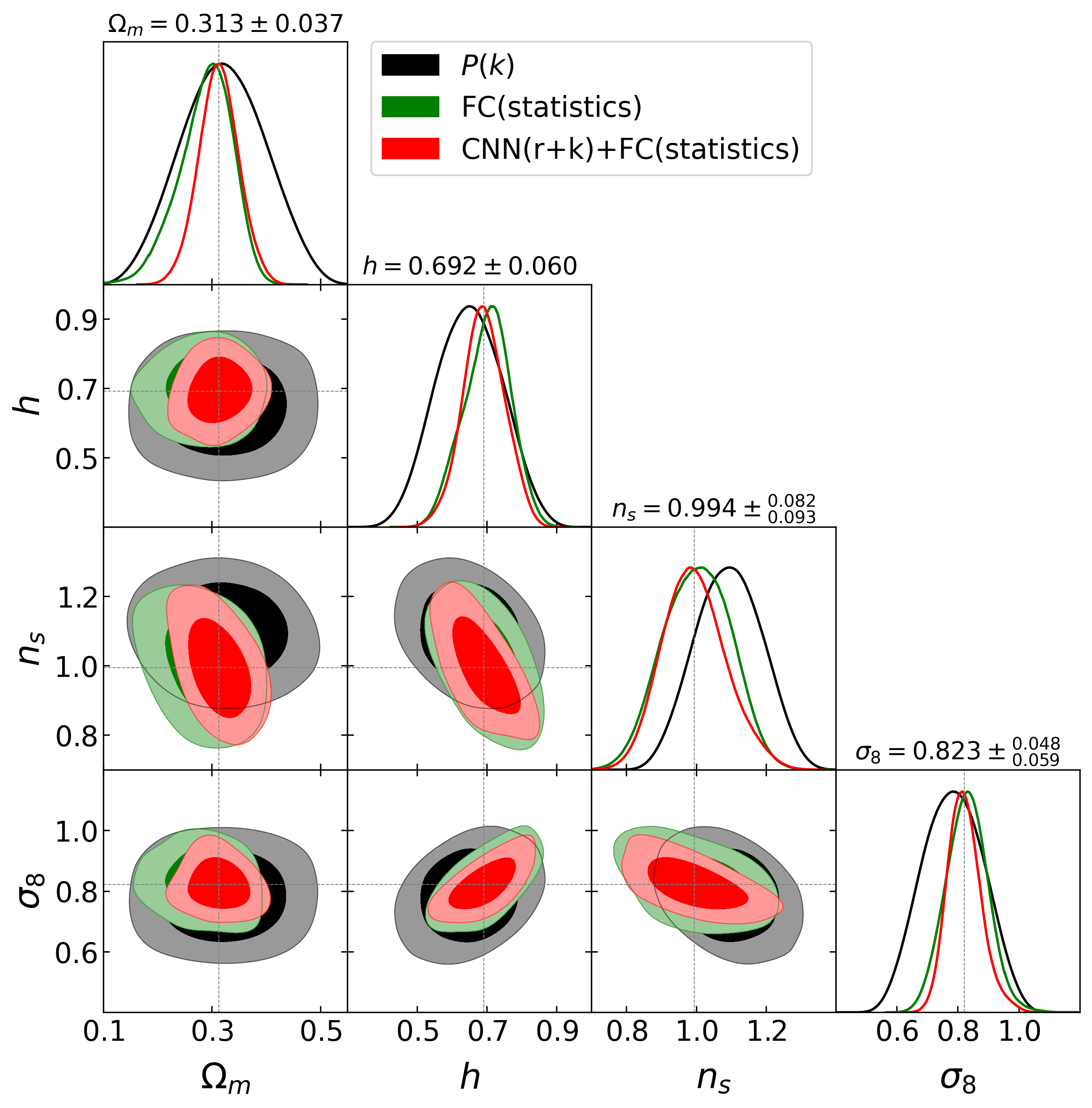}
	\caption{Probability distributions of the  cosmological parameters predicted from the model ``CNN($r+k$)+FC(statistics)'' in lCNN (red) and ``FC(statistics)''  (green). The averaged value (gray dashed) of each parameter over all test datasets is displayed for comparison. { Note that the 1D and 2D distributions are calculated directly from the 1D histrogram and the joint distribution of the bias (as defined in Eq.~\ref{eq:deltap}) for the parameters. The results of the likelihood-based analysis performed on the power-spectrum statistics alone, $P(k)$, are shown in black in the 1D and 2D distribution plots.}}
	\label{fig:deltap}
\end{figure*}

To investigate the {error} of lCNN in prediction, we illustrate the joint distribution of each parameter pair and the histogram of each parameter in Fig.~\ref{fig:deltap}. Two models are presented for comparison: ``CNN($r+k$)+FC(statistics)'' and {``FC(statistics)''}.For clarity, the plots display the distributions of the errors of cosmological parameters, centered around the mean of the parameter space, i.e.
\begin{equation}\label{eq:deltap}
p_i = \Delta p_i +\bar{p}^{\rm true}\,,~\quad {\rm with~} \Delta p_i =  p_i^{\rm pred} - p_i^{\rm true}
\end{equation}
where $\bar{p}^{\rm true}$ denotes the averaged true value over all 210 test samples with varied cosmological parameters. Here $\Delta p_i$  denotes the {\it bias} for a given parameter predicted from the $i$-th test sample.

\begin{figure}[htpb]
	\includegraphics[width=\columnwidth]{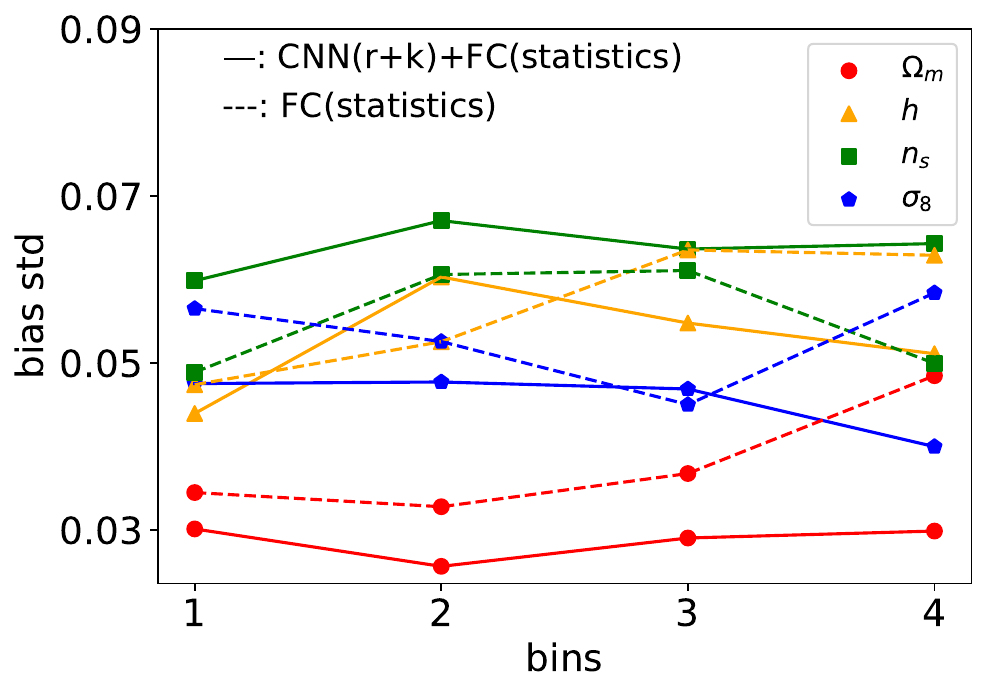}
	\caption{{ Estimate of the standard deviation of bias (defined as $\mathrm{bias} = p_i^{\rm pred} - p_i^{\rm true}$) for different parameter ranges. Each parameter value was linearly divided into four bins. Based on the range of values in each bin, we selected test samples and then computed the standard deviation of the bias from the lCNN prediction. Solid lines represent the model ``CNN($r+k$)+FC(statistics)'', dashed lines represent the model ``FC (statistics)'', and different colors denote different parameters. The standard deviations of the bias show slight variations across parameters.}
 }
	\label{fig:errorbin}
\end{figure}

{ In contrast to likelihood analysis, error estimation in machine learning requires a careful consideration.~\cite{Jeffrey2020SolvingHP} suggests defining parameter error using marginal flows and Moment Networks. ~\cite{2018arXiv180605978S} introduces an uncertainty estimation method using Bayesian convolutional neural networks with variational inference. Another approach~\citep{2023RAA....23g5011Z}, involves predicting error by sampling hidden variables. Alternatively, prediction accuracy~\citep{lazanu2021extracting} can provide an error estimation. In this study, we use a large sample of true parameter values. The distribution of the bias, $\Delta p_i$, between the predicted and true values as defined in Eq.~\ref{eq:deltap}, yields a probability distribution of biases in parameter accuracy, which directly corresponds to our neural network's error estimation. To test the robustness of these error estimates, we divided the test sample into four bins based on parameter values, as shown in Fig.~\ref{fig:errorbin}. This approach allows us to estimate the bias distribution within specific parameter ranges and calculate the standard deviation, which serves as the error for each parameter range. Theoretically, using more bins increases the accuracy of the error estimate, and the standard deviation of the bias more closely approximates the true error. As shown in Fig.~\ref{fig:errorbin}, the standard deviation of the bias distribution does not vary significantly with different parameter values. Additionally, these standard deviation values are essentially the same as the error values in Fig.~\ref{fig:deltap}, indicating that our error estimation is reasonable.}

{ To further justify the validity of using this method for error estimation, we conducted a likelihood-based analysis on the statistics of the power spectrum alone. Initially, we computed the covariance matrix of $P(k)$ using 1000 Quijote realizations of the fiducial cosmology in the range of  $k\in[0.05, 0.6] ~h\mathrm{Mpc}^{-1}$. The likelihood can be constructed through $-2\ln\mathcal{L}(P(k)|\bm{\theta}) \propto [P(k,\bm{\theta})- \bar{P}(k)]^T\Sigma^{-1}[P(k,\bm{\theta})- \bar{P}(k)]$, where $\bar{P}(k)$ is the ``true'' power, estimated from the mean spectrum of all mock realizations. The covariance $\Sigma$ is primarily sourced from the sampling variance of $P(k)$ in the mock data and was directly estimated from the Quijote fiducial-cosmology mock realizations.}

{ 
We performed our likelihood evaluation using the Monte Carlo Markov Chain (MCMC) method, utilizing the emcee package~\citep{2013PASP..125..306F}. In the MCMC process, the $P(k,\bm{\theta})$ data for a given cosmological parameter set were derived from the 1710 cosmologies by performing the nearest interpolation in high-dimensional space, which enables us to estimate $P(k,\bm{\theta})$ for any sampled point in the parameter space. The cosmological constraints from $P(k)$ alone, based on the likelihood inference, are summarized in Fig.~\ref{fig:deltap}. As seen, the $P(k)$-alone-derived 1$\sigma$ statistical errors are 0.069 for $\Omega_m$, 0.085 for $h$, 0.090 for $n_s$, and 0.091 for $\sigma_8$. Thus, the constraints are weaker than those from lCNN, especially for $\Omega_m$ and $\sigma_8$.
}

The two-dimensional contour plots illustrate the joint probability distribution at 68\% and 95\% levels, respectively, providing information about the correlation between each pair of parameters. Meanwhile, the one-dimensional distribution displays the marginalized probability of each parameter. As observed, for each parameter, the model ``CNN ($r+k$)+ FC(statistics)'' offers a more sharply marginalized probability distribution than that for ``FC(statistics)''. In other words, the former provides more accurate predictions on the cosmological parameters. This finding is further confirmed by the contour plots.

The contour area corresponds to the statistical uncertainty. As observed, the scatters of the difference between the predicted values and the true ones for all test datasets, derived from lCNN are considerably smaller than those from the case ``FC(statistics)''. Both the centers of the two-dimensional contour and the one-dimensional distribution for each parameter are close to the averaged true value (gray dashed), and the deviation is significantly smaller than the statistical uncertainty. In particular, the mean and the associated standard deviation, $\sigma_p$, derived from the model ``CNN(r+k)+FC(statistics)'' for the marginalized distribution are listed at the top of each one-dimensional plot. As observed, all the mean values closely agree with the true ones.  The deviation for ``CNN($r+k$)+FC(statistics)'' from the averaged true value is 2.8\% for $\Omega_m$, 2.4\% for $h$, 2.5\% for $n_s$, and 0.6\% for $\sigma_8$, respectively.  In comparison with using the model ``FC(statistics)'', lCNN yields smaller statistical errors. { Specifically, it is reduced by 23\% for $\Omega_m$, 11\% for $h$, 8\% for $n_s$ and 21\% for $\sigma_8$, respectively. In comparison with using the likelihood-based analysis on $P(k)$ data, our lCNN performs much tighter constraints on parameters, especially on $\Omega_m$ and $\sigma_8$.}

We also observe that there is almost no correlation between the predicted parameters overall, although weak correlations exist between certain parameter pairs, such as $n_s$ and $h$. Since the parameters in the test sample were randomly generated and uncorrelated, there should be no significant correlation between the values of the parameters predicted by ICNN. The results in the test sample meet our expectations, demonstrating that our ICNN model can achieve high accuracy in parameter prediction with statistical errors that are smaller than that of the conventional likelihood analysis based on the power-spectrum statistics alone.

\section{Concluding Remarks}\label{sect:5}
In this study, we have designed a lightweight deep convolutional neural network, lCNN, aimed at estimating cosmological parameters from simulated three-dimensional DM halo number density field  and associated statistics. Our training dataset consists of 2000 realizations of a cubic box with a side length of 1000 $h^{-1}{\rm Mpc}$, each sampled with $512^3$ DM particles and $512^3$ neutrinos interpolated over a cubic grid of $300^3$ voxels. Under the flat $\Lambda$CDM model, simulations vary the standard six cosmological parameters, including $\Omega_m$, $\Omega_b$, $h$, $n_s$, $\sigma_8$, $w$, along with the neutrino mass sum, $M_\nu$. 

{Seven} distinct models have been considered to assess the optimal input datasets, including: ``CNN($r$)'', which utilizes solely the density field; ``CNN($r+k$)'', incorporating both density field and its Fourier modes; ``CNN($r$)+FC(statistics)'', employing the density field along with three statistics (i.e., the halo density power spectrum, the 2PCF, and the WST coefficients); ``CNN($r+k$)+FC(statistics)'', combining ``CNN($r+k$)'' with three statistics; ``RF(statistics)'', utilizing the random forest neural network trained solely with the three measured statistics, for comparison with the lCNN;{``FC(statistics)'' and FC(Pk), utilizing the fully connected neural network trained solely with the three measured statistics and $P(k)$ respectively}.

Our findings reveal several key insights: 1) within the framework of lCNN, extracting LSS information is more efficient from the halo density field compared to relying on statistical quantities including the power spectrum, 2PCF, and WST coefficients; 2) combining the halo density field with its Fourier-transformed counterpart enhances predictions, and augmenting the training dataset with measured statistics further improves performance; 3) the neural network model achieves high accuracy in inferring $\Omega_m$, $h$, and $\sigma_8$, while showing inefficiency in predicting $\Omega_b$, {$n_s$}, $M_\nu$, and $w$; 4) moreover, compared to the simple fully connected network trained with three statistical quantities, our proposed lCNN model yields high prediction accuracy in the parameters and provides smaller statistical errors, {reducing the errors by about 23\% for $\Omega_m$, 11\% for $h$, 8\% for $n_s$, and 21\% for $\sigma_8$, respectively; 5) Compared to the likelihood-based analysis of the $P(k)$ data, our lCNN achieves significantly tighter constraints on parameters, particularly on $\Omega_m$, $h$, and $\sigma_8$, reducing them by 46\%, 29\%, and 40\%, respectively.} 

{ Note that the previous constraints~\citep{villaescusa_Navarro_2020, Massara2020UsingTM}, especially for the neutrino mass sum, $M_{\nu}$ are considerably tighter than those derived here. This is because their estimations are optimal for two primary reasons: i) they are derived from the total matter density power spectrum, which cannot be directly or accurately measured from real observations. In contrast, our observable is the spatial distribution of halos and the associated statistics. 2) To more accurately reflect real observations, we have incorporated the effects of RSD, coordinate mapping from fiducial cosmology, and fixed the halo number density in our dataset. These additional observational effects included in our mock data have the potential to significantly weaken the cosmological parameter constraints. }

Machine learning is highly effective at analyzing complex features in complicated datasets. From this perspective, a limitation of our study is that our training samples are composed of sparse halo fields with a low number density of \(2 \times 10^{-4}\). Consequently, many small-scale structures and clustering details are not captured in such sparse fields. A promising direction for future investigation would be to increase the number density by one to two orders of magnitude to better mimic the observational data from stage-IV surveys. In such scenarios, we expect that machine learning could significantly enhance performance and offer substantial advantages over traditional statistical methods.

In future work, we intend to evaluate the ability of the network to predict cosmological parameters from light-cone simulations, and finally, apply it to real observational data.

\begin{acknowledgments}
We thank Francisco Villaescusa-Navarro and Yin Li for helpful discussions. This work is supported by National SKA Program of China (2020SKA0110401, 2020SKA0110402, 2020SKA0110100), the National Key R\&D Program of China (2020YFC2201600, 2018YFA0404504, 2018YFA0404601), the National Science Foundation of China (11890691, 12203107, 12073088, 12373005), the China Manned Space Project with No. CMS-CSST-2021 (A02, A03, B01), the Guangdong Basic and Applied Basic Research Foundation (2019A1515111098), and the 111 project of the Ministry of Education No. B20019. We also wish to acknowledge the Beijing Super Cloud Center (BSCC) and Beijing Beilong Super Cloud Computing Co., Ltd (\url{http://www.blsc.cn/}) for providing HPC resources that have significantly contributed to the research results presented in this paper.
\end{acknowledgments}

\nocite{*}

\bibliography{main_apssamp}

\appendix
\section{Validating the robustness of lCNN}\label{sec:appendixA}

{ In order to validate the robustness of these results with regard to architectural choices, we proceeded to train nine additional CNNs with varying numbers of convolutional and fully connected layers using only the density field, i.e., ``CNN($r$)''. For these networks, we employed a $(3,3,3)$ convolutional kernel with zero padding to ensure that the original output data size was preserved when integrating convolutional layers into our baseline network. We ensure that the newly added convolutional layers maintain consistency in both the input and output feature numbers. 

Furthermore, fully-connected layers were added before the output layer of the base network. Each model was trained for 1000 epochs with identical hyperparameters, after which their performance on the test data was evaluated to determine the epoch at which overfitting occurred. 

As illustrated in Tab.~\ref{tab:netrobust}, the various models exhibit overfitting at approximately the same number of epochs, and their losses are also approximately equivalent. Nevertheless, an increase in the loss function is observed when the fully connected (FC) layer reaches six layers. This phenomenon is due to the fact that the incorporation of numerous fully connected layers results in a considerable increase in the number of parameters, thereby rendering the model more challenging to converge. Consequently, the findings indicate that, within a specific range, the model is capable of adapting to variations in its architectural design.}

\begin{table*}[htpb]
\centering
\begin{adjustbox}{width=1.7\columnwidth,center}
\begin{tabular}{c|c|c|c|c}
\hline
\hline
\diagbox{Conv-layers}{Overfitting epoch/minimum loss}{FC-layers} & 3 & 4 & 5 & 6 \\
\hline
4 &\bf{880/0.77} &- &- &- \\
\hline
5 & -& 800/0.76&875/0.80 & 845/0.81\\
\hline
6 & -&935/0.73 &975/0.79 &820/0.81 \\
\hline
7 &- &950/0.75 & 975/0.76&965/0.80 \\
\hline
\end{tabular}
\end{adjustbox}
\caption{{ Validation of the robustness of lCNN concerning architectural choices. Nine additional CNNs were trained with varying numbers of convolutional and fully connected layers to observe changes in overfitting epochs and minimum loss, using only the density field, i.e., ``CNN($r$)'' as a test. The bold text represents the results of our benchmark network, which has 4 convolutional layers and 3 fully connected layers.}}
\label{tab:netrobust}
\end{table*}

\section{Validating the robustness of DM halo data selection }\label{sec:appendixB}

\begin{figure}[htpb]
	\includegraphics[width=0.9\columnwidth]{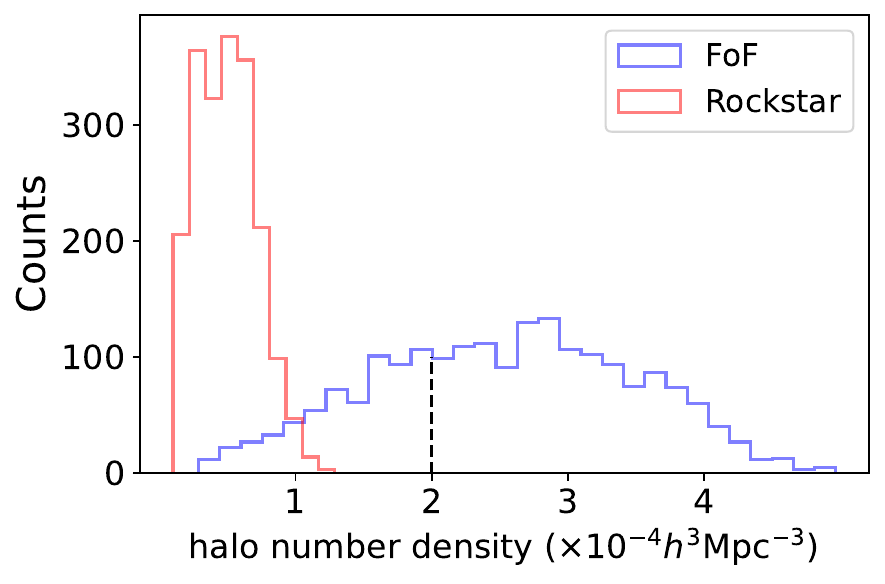}
    \centering
	\caption{{ Histograms of the halo number density for the Rockstar halo catalogs (red) and the FoF halo catalogs (blue) of the Quijote simulations, where the FoF catalogs have been applied with a minimum mass cutoff of  $M>M_{\rm min}$ with $M_{\rm min}= 10^{13} M_\odot h^{-1}$. The dotted line represents the halo number density of $n_{\rm fix} = 2 \times 10^{-4}~h^3{\rm Mpc}^{-3}$, which is consistent with the current spectral observations. An increase in $M_{\rm min}$ allows for a reduction in the number densities of the FoF catalogs (to the right of the dotted line) to $n_{\rm fix}$.}        }
	\label{fig:rockn}
\end{figure}

\begin{figure*}[htpb]
	\includegraphics[width=1.7\columnwidth]{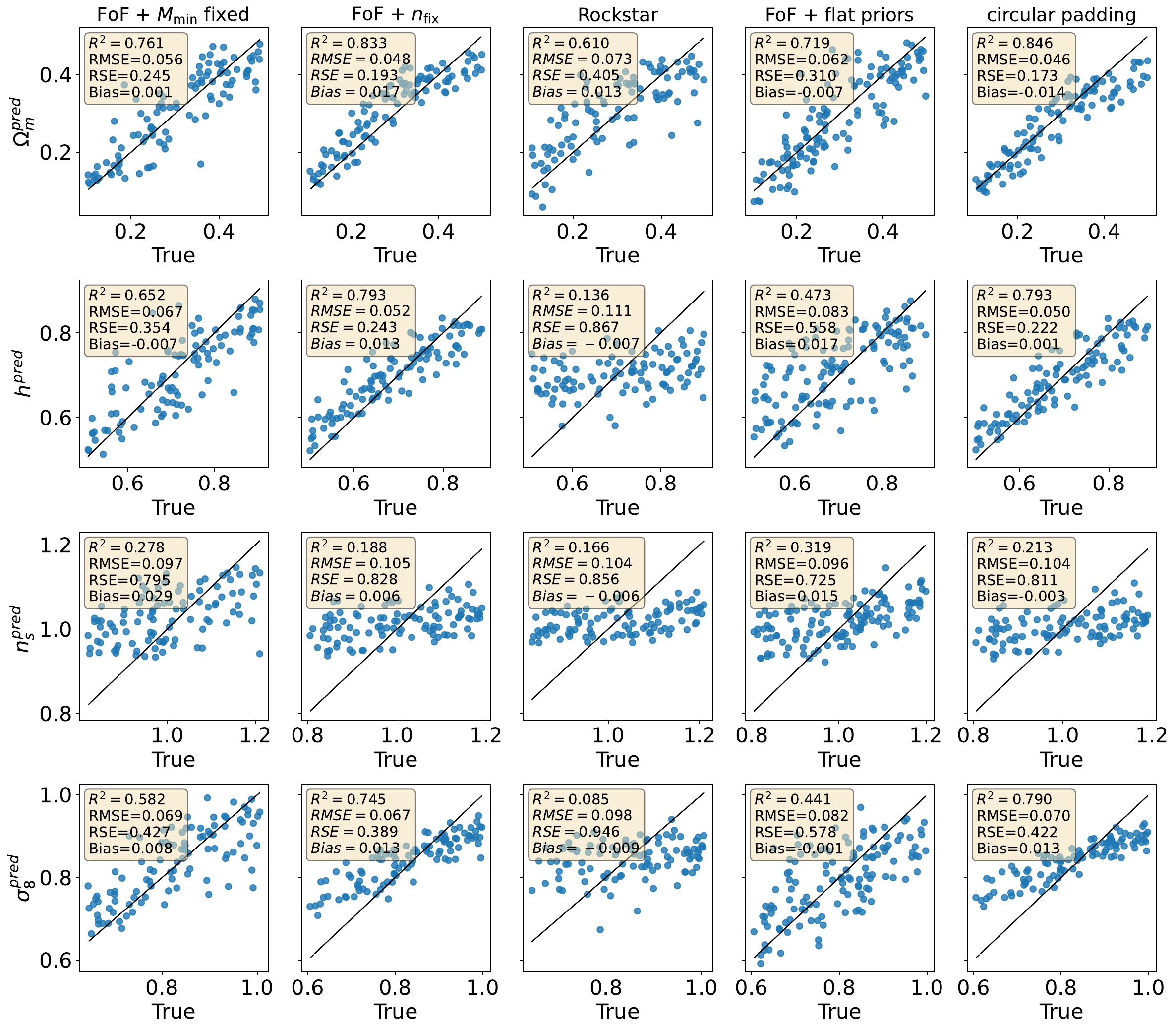}
    \centering
	\caption{
 { Comparison between the true values of $(\Omega_m, h, n_s, \sigma_8)$ (from top to bottom) from the test sample and their predictions from models trained with different  DM halo datasets. The black lines indicate perfect prediction matching the truth. The resulting $R^2,\mathrm{RMSE},\mathrm{RSE},\mathrm{BIAS}$ values for each case are also shown. The data employed in the first to fourth columns, respectively, are as follows: the fixed halo minimum mass cutoff of $M_{\rm min} = 10^{13}M_\odot h^{-1}$, the fixed halo number density of $n_{\rm fix} = 2\times 10^{-4}~ h^3(\mathrm{Mpc})^{-3}$ from the FoF halo catalogs (the fiducial case), the Rockstar halo catalogs, and the FoF halo catalogs, but recovering flat priors of $\sigma_8$ and $n_s$ through data augmentation.The fifth column and the second column use the same data and data processing methods, the only difference is that the second column uses zero padding, while the fifth column uses circular padding.}
 }
\label{fig:rock}
\end{figure*}

{
In the following, we will focus on the statistical analysis of ``CNN($r$)'' as an illustrative case to facilitate a quantitative comparison. A discussion of the impact of DM halo data selection on the results of our study was carried out with the following three halo datasets explored:
\begin{enumerate}
\item ``FoF+$M_{\rm min}$ fixed'' -- replacing the fixed number density with a fixed minimum DM halo mass cutoff of $M_{\rm min} = 10^{13} M_\odot h^{-1}$ to generate FoF catalogs as training data.

\item ``FoF+$n_{\rm fix}$'' -- the fiduical catalogs generated by fixing the halo number density to $n_{\rm fix} = 2 \times 10^{-4}~h^3{\rm Mpc}^{-3}$ for the FoF catalogs. 

\item ``Rockstar'' -- using the Rockstar halo catalogs from the Quijote simulations instead of the FoF halo catalogs.

\item ``FoF+flat priors'' -- recovering $\sigma_8$ and $n_s$ to flat priors through data augmentation by various means of reflecting and rotating the original halo density field.
\end{enumerate}

After training on each of the datasets, the performance of the parameter recovery was evaluated based on each corresponding test dataset, as shown in Fig.~\ref{fig:rock}.  

The first column of Fig.~\ref{fig:rock} represents the results for the FoF catalog with a minimum mass of $10^{13}~M_\odot/h$ fixed. This adjustment flattens the priors of parameters such as $\sigma_8$ in the training data, leading to improved performance in parameter spaces with sparse training data. Consequently, the reconstruction of $\sigma_8$ at lower values shows better accuracy compared to the fiducial case. Additionally, the reconstructions for the other parameters remain comparable to those in the fiducial one, indicating that the results are not sensitive to the choice between $n_{\rm fix}$ and $M_{\rm min}$.

In comparison to our fiducial case (the second column) and the other two cases, Rockstar halo catalogs result in larger RMSE values across all listed cosmological parameters. This is due to the fact that each Rockstar halo catalog typically has a significantly lower halo number density than the other catalogs, which results in a reduction in the amount of cosmological information encoded.

In comparison to Fig.~\ref{fig:para_dis}, it is evident that the fiducial halo catalogs result in the distribution of $\sigma_8$ being absent at low values. Consequently, following the recovery of the flat prior on $\sigma_8$ in the case of ``FoF+flat prior'', the augmented training data can provide additional information, thereby reducing the bias value and more accurately predicting $\sigma_8<0.8$ than in the other cases.  
}

{In our work, we applied zero padding in the first two convolutional layers of the CNN (please refer to the second column of Fig.~\ref{fig:rock} for the corresponding results). For comparison, we discussed circular padding (see the fifth column of Fig.~\ref{fig:rock}), which is suitable for data with periodic boundary conditions. As observed, circular padding provides a slight improvement in training performance over zero padding, though the results are generally similar. The discrepancy between the two padding methods is not substantial, as evidenced by the fact that 97.7\% to 98.1\% of the pixel values in the 1710 samples used are zero.}

\end{document}